\documentclass[hidelinks, runningheads]{llncs}
\usepackage[T1]{fontenc}
\usepackage{multirow}
\usepackage{hyperref}
\usepackage{booktabs}
\usepackage{comment}
\usepackage{makecell}
\usepackage{soul}
\usepackage{xcolor}

\usepackage{multirow}
\usepackage{graphicx}

\usepackage{amsmath}
\begin{document}
\title{A multi-channel cycleGAN for CBCT to CT synthesis}
%
%
\author{ Chelsea A. H. Sargeant\orcidID{0000-0002-8737-4835} \and Edward G. A. Henderson\orcidID{0000-0003-3752-4054} \and D\'{o}nal M. McSweeney\orcidID{0000-0002-7099-8270}  \and Aaron G. Rankin\orcidID{0000-0002-0386-1119} \and
Denis Page\orcidID{0009-0005-9015-180X}\thanks{Authors listed in order of contribution.}}

\authorrunning{C. Sargeant \textit{et al.}}
%
\institute{Division of Cancer Sciences, The University of Manchester
\\
\email{chelsea.sargeant@postgrad.manchester.ac.uk}}
\maketitle              
\begin{abstract}
Image synthesis is used to generate synthetic CTs (sCTs) from on-treatment cone-beam CTs (CBCTs) with a view to improving image quality and enabling accurate dose computation to facilitate a CBCT-based adaptive radiotherapy workflow. As this area of research gains momentum, developments in sCT generation methods are difficult to compare due to the lack of large public datasets and sizeable variation in training procedure. To compare and assess the latest advancements in sCT generation, the \textit{SynthRAD2023} challenge provides a public dataset and evaluation framework for both MR and CBCT to sCT synthesis.
Our contribution focuses on the second task, CBCT-to-sCT synthesis. By leveraging a multi-channel input to emphasize specific image features, our approach effectively addresses some of the challenges inherent in CBCT imaging, whilst restoring contrast necessary for accurate visualisation of patients anatomy. Additionally, we introduce an auxiliary fusion network to further enhance the fidelity of generated sCT images.

\keywords{synthetic CT  \and multi-channel cycleGAN \and CBCT quality}
\end{abstract}
\section{Introduction}

In photon and proton radiotherapy, on-treatment cone-beam computed tomography (CBCT) imaging plays a crucial role in detecting anatomical changes and ensuring accurate patient positioning for precise treatment setup \cite{Oldham2005}. However, CBCT image quality often suffers from artefacts such as scatter, noise and beam hardening, resulting in inferior image quality compared to planning CT (pCT) scans. These artefacts lead to inaccuracies in Hounsfield unit (HU) values, rendering CBCT images unsuitable for accurate dose calculation.

To address these challenges, deep learning-based image synthesis techniques have been successfully employed to enhance CBCT image quality to a level comparable to CT scans. This involves using neural networks to discover a complex mapping between source (CBCT) and target (CT) image domains, enabling the generation of synthetic CT (sCT) images directly from CBCT data \cite{Rusanov2022}. The primary objective is to produce sCT images that retain the anatomical structure of the CBCT scans while closely approximating the HU of the pCT images to facilitate accurate dose calculation using on-treatment imaging.

Cycle-consistent generative adversarial networks (cycleGANs) have emerged as the dominant methodology for the conversion of CBCT to CT images for various anatomical sites \cite{Harms2019,Kurz2019,Maspero2020}. CycleGANs use two generator-discriminator network pairs, trained concurrently, to allow the model to learn the translation function from unpaired images \cite{Zhu2017}. The generative networks are responsible for converting between the CBCT and CT image domains. One generator produces CT-like output (a sCT) from CBCT input, while the other performs the reverse conversion. The discriminators learn to differentiate between real and synthetic images. Once trained, a single generator is employed for the desired translation, in this context, the CBCT-to-CT generator.


Our approach is an extension of the 2D cycleGAN architecture originally proposed by Zhu \textit{et al.} \cite{Zhu2017}. The generator architecture has been modified to incorporate long-range skip connections and self-attention layers to improve model performance. Crucially, we separate the input CBCT and CT images into multiple channels, each emphasizing specific image features, and introduce an auxiliary fusion network to recombine the multi-channel predictions. The incorporation of multi-channel information and subsequent fusion of these channels accounts for the diverse variety of input images and contributes to the enhanced fidelity of generated sCT images. 
Our method is robust to the challenges of CBCT imaging. It effectively mitigates the impact of significant scatter artefacts often encountered in CBCT scans, whilst successfully restoring contrast necessary for accurate visualisation of the soft tissues within the patients. This approach enables the generation of high-quality synthetic CT images from CBCT scans. With these extensions, we aim to produce sCT images of high enough quality to facilitate accurate dose calculations on data acquired from on-treatment CBCT imaging.

\section{Method}
\subsection{Dataset}

Imaging data of patients who underwent brain or pelvis radiotherapy were collected from three Dutch institutions: Radboud University Medical Center, University Medical Center Utrecht and University Medical Center Groningen for the \textit{SynthRAD2023} challenge. Here, we focused on the second of two tasks, which involved generating synthetic CTs from CBCTs. The training, validation and testing cohorts were determined prior to the challenge by the organisers, more details can be found in Table 1 in Thummerer \textit{et al.} \cite{Thummerer2023}. A total of 360 CBCT-CT pairs were available, split evenly across the brain and pelvis. \par 
All pre and post-processing was applied to 3D images, however, individual axial slices were extracted from all image volumes (CT and CBCT) for training. A random 80/20 split was used to create train-time training and validation sets. In our case, two separate site-specific models were trained: one for brain scans and another for pelvis scans. Training and validation splits for each site categorised by center name (A, B, C; as in \cite{Thummerer2023}) are shown in Table \ref{tab:dataset}. For image acquisition parameters, see \cite{Thummerer2023}.

\begin{table*}[!htbp]
\caption{Train-time training and validation splits. Values represent number of axial slices from each center.}
\label{tab:dataset}
\centering
\setlength\tabcolsep{1.2mm}
\begin{tabular}{l cccc  ccccc }
\toprule 
& \multicolumn{4}{c}{\textit{Brain}} & \hspace{2cm}& \multicolumn{4}{c}{\textit{Pelvis}}\\
& A & B & C & \textbf{Total} & & A & B & C & \textbf{Total} \\
\midrule
\textbf{Training} & 10,955& 8,262& 10,434 & 29,651 & & 4,772 & 4,111 & 2,369 & 11,252\\
\vspace{1pt}\\
\textbf{Validation} & 2,711&2,012&2,690&7,413 &  & 1,194 & 1,056 & 563 & 2,813 \\
\midrule
\textbf{Total} & 13,666 & 10,274 & 13,124 & 37,064 & & 5,916 & 5,167 & 2932 & 14,065 \\
\bottomrule
\end{tabular}
\end{table*}

\subsection{Pre-Processing}\label{sec:preprocessing}
Preprocessing was performed by the challenge organisers resulting in 3D CT and CBCT images with corresponding binary masks of the patient. All images of an anatomical region were resampled to the same voxel spacing: $1 \times 1 \times 1 \text{mm}^3 $ and $1 \times 1 \times 2.5 \text{mm}^3 $ for the brain and pelvis, respectively. A full description of the preprocessing can be found in \cite{Thummerer2023}. 

Following the initial preprocessing, images were processed as follows:
\begin{itemize}
\item Correction of masks and CBCT image range;
\item Overflow correction for pelvis patients;
\item Mask application;
\item Multi-channel range selection and normalisation;
\item Padding/Cropping.
\end{itemize}

These preprocessing steps were also applied on inference (with the exception of the mask correction). The challenge dataset was curated to represent the variation in a realistic multi-centre setting; this included variation in image protocol, CBCT quality, field-of-view (FOV) and severity of artefacts. To capture as much information as possible, we opted to include all available patients in the training set and handle the variability of the dataset with a multi-channel approach. 
Below, we present a more detailed description of this pipeline.

\subsubsection{Mask and image range correction} 
All provided images and masks were visually inspected to ensure consistency across the training cohort. For a large number of pelvis cases, the masks included large abnormalities resulting in missing regions within patient body and were deemed unsuitable for training. As a result, the masks were regenerated using thresholding and binary dilation and morphological closing functions from \textit{SimpleITK}. The CBCTs from centre A had to be shifted by $-1024$ HU to ensure the CBCTs from all centres were contained within a consistent range of $[-1024, 3000]$ HU.

\subsubsection{Overflow correction}
Upon visual inspection it was also noticed that CBCTs from $55/60$ pelvis patients from Centre C suffered from high intensity artefacts on the surface of the patient that were not seen in the corresponding CT. We believed this to be due to a numerical overflow error occurring during the CBCT image reconstruction. 
This was corrected prior to training by overwriting the artefact voxels with air (-1024 HU). In detail, a distance transform was used to create a $\sim 40\text{mm}$ thick hull around the external contour of the patient to capture high-intensity values (\textgreater 1000 HU) resulting from the overflow error. This correction was applied to all pelvis patients since centre identity was unavailable on testing.

\subsubsection{Multi-channel input}
\begin{figure}[h!]
    \centering
    \includegraphics[width=\textwidth]{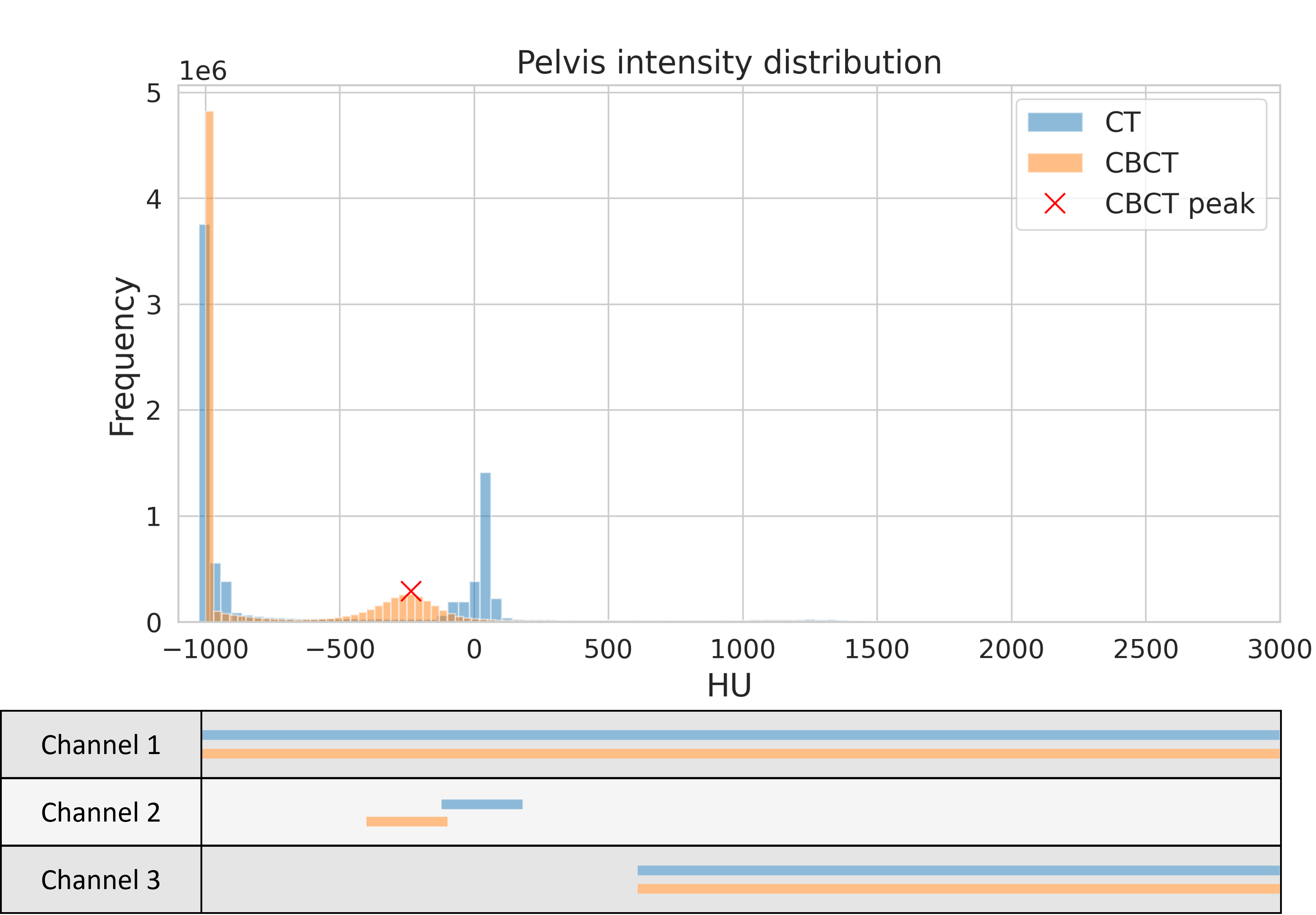}
    \caption{CBCT and CT image histogram for a pelvis scan. Below the histogram shows the ranges of each of the three input channels. Channel one accounts for the entire image. Channel two captures soft tissue information by centering the channel on either the water peak in the case of the CT image or on a peak found on a per-patient basis on the CBCT. In either case a window of $\pm 150$ was used. High-density structures, such as certain bones or artificial metal implants, were accounted for in the third channel.}
    \label{fig:HM-channels}
\end{figure}

\begin{figure}[ht!]
    \centering
    \includegraphics[width=11cm]{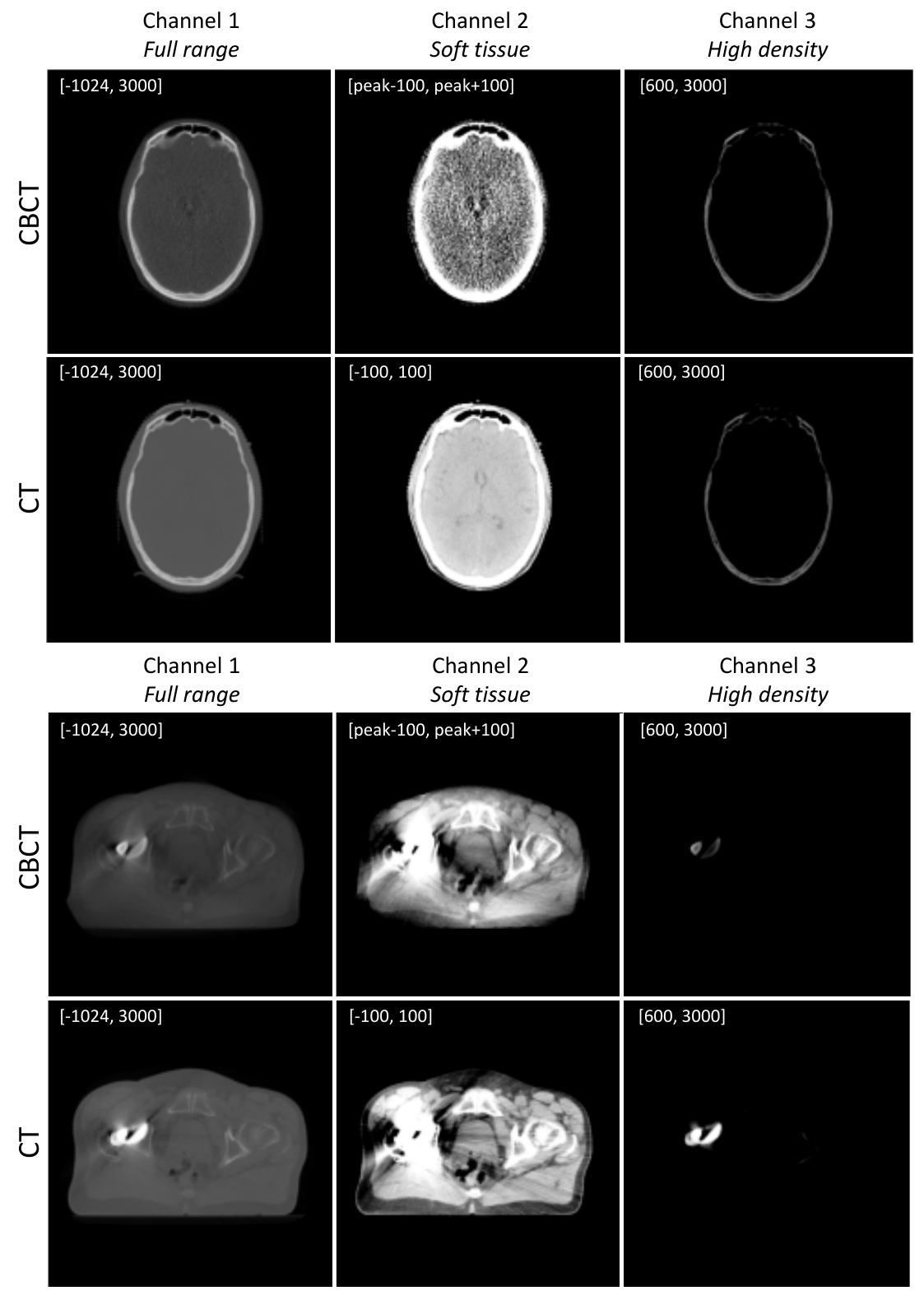}
    \caption{Three channel input for the brain and pelvis model. Channel one captures the full image window, channel two is the soft tissue window and channel three captures high-density structures such as high-density bone in the brain patient and metal implants in the pelvis patient displayed. The range is reported in HU, and each channel is independently normalised to [0, 1].}
    \label{fig:3channel-input}
\end{figure}

The CT and CBCT scans were normalised in three separate channels using windowing to enhance the contrast of structures within the patient. In the first channel, we use the full width of the image range $[-1024, 3000]$ HU. In channel two, we use a contrast setting used to view soft tissue structures within the region of interest; for the CT this was $[-150, 150]$ HU and $[-100, 100]$ HU for the pelvis and brain, respectively, chosen to be centred on the water peak.
Due to a lack of calibration of the CBCT images, both within and among centres, an automated peak finder was implemented. For this we used the $find\_peaks$ function from the \textit{SciPy} package. The window width remained the same; either $\pm 150$ HU and $\pm 100$ HU depending on the image site, however, the level varied on a patient basis depending on the peak found within the image intensity histogram as shown in Figure \ref{fig:HM-channels}.
In the final channel, both the CT and CBCT images were clipped to $[600, 3000]$ HU to capture information about the high-density structures within the patient: high-density bone and any metal present due to artificial hips, pins, and dental work, for example. Each channel was then independently scaled to the range $[0, 1]$ using min-max normalisation to improve training stability. Figure \ref{fig:3channel-input} depicts the three-channel input for both the CBCT and CT images following range selection and min-max normalisation.

\subsubsection{Cropping}
The three-channel images were padded or cropped to height and width of $448\times448$ voxels for the pelvis and $304\times304$ voxels for the brain.

\subsubsection{Model input}
 The model input consisted of three-channel CT/CBCT axial slices, paired only by rigid registration.
 
\subsection{Proposed Method} 
Our method is based on a traditional cycleGAN architecture \cite{Zhu2017} with some specific modifications for this challenge. CycleGANs consist of two sets of competing generator and discriminator networks trained simultaneously to enable unpaired training. The generators learn to produce CT-like output from CBCT input (a sCT) and vice versa (synthetic CBCT (sCBCT) output from CT input). The discriminators learn to classify CT/sCT and CBCT/sCBCT input as real and fake.
Additionally, cycle consistency was enforced by inputting the synthetic image into the reverse generator to reverse the translation back to the original image modality.

\subsubsection{Disciminator network}
The discriminator used in our modified cycleGAN was standard and directly adapted from the PatchGAN \cite{Isola2017}. 
The total loss for the discriminators is
\begin{equation}
    \mathcal{L}_D = \mathcal{L}_{D_{CT}} + \mathcal{L}_{D_{CBCT}} 
\end{equation}
where ${D_{CT}}$ is the discriminator aiming to classify sCT from pCT and ${D_{CBCT}}$ is the discriminator aiming to classify sCBCT from CBCT. The loss function used was binary cross-entropy.  

\subsubsection{Generator network}
The generator architecture used for our cycleGAN was based on the convolutional neural network detailed by Zhu \textit{et al.} \cite{Zhu2017}. This has three convolutional layers with pooling, followed by nine residual convolutional blocks and then three resize convolutional layers, described by Odena \textit{et al.}, to re-establish the original spatial dimensions \cite{Odena2016}. We added UNet-like long-range skip connections between corresponding down- and up-sampling convolutional levels to preserve contextual information \cite{Ronneberger2015}. Finally, we added attention gates to the skip connections to emphasize salient features propagated forward from earlier in the network \cite{Schlemper2019}.

The generator receives three-channel input and produces three-channel output. This is performed for ease of implementation of the cycle consistency aspect, where the output of each generator is also sent through the reverse generator.

\subsubsection{Fusion network}
An auxiliary fusion network was added onto the cycleGAN to fuse the output three channels from the CBCT-sCT generator into a single-channel greyscale image. This final network had identical architecture to the generators but contained only a single residual block, and short-range residual connections across all convolutional layers as in Milletari \textit{et al.} to speed up convergence \cite{Milletari2016}.

A mean square error loss term was added to train the fusion network to maximise the similarity between the output single-channel image and the CT.
\\
\newline
\noindent The total loss for the generators is
\begin{eqnarray}\label{eqn:gen_loss}
    \mathcal{L}_{G} & = & \mathcal{L}_{G_{\text{GAN-sCT}}} + \mathcal{L}_{G_{\text{GAN-sCBCT}}} \nonumber \\
                    &   & + \alpha (\mathcal{L}_{G_{\text{cycle-CT}}} +\mathcal{L}_{G_{\text{cycle-CBCT}}}) \\
                    &   & + \beta  (\mathcal{L}_{G_{\text{identity-CT}}} +\mathcal{L}_{G_{\text{identity-CBCT}}}) \nonumber \\
                    &   & + \mathcal{L}_{G_{\text{fusion-sCT}}} \nonumber
\end{eqnarray}

where $\alpha = 10$ and $\beta = 5$. Each generator loss was calculated using the mean squared error. 
The $\mathcal{L}_{G_{\text{GAN}}}$ loss terms reward the generator for deceiving the discriminator.
The $\mathcal{L}_{G_{\text{cycle}}}$ loss terms, or consistency loss, are performed between CT and sCT (or CBCT and sCBCT) pairs where the synthetic image has been passed through both generators in a cycle.
$\mathcal{L}_{G_{\text{identity}}}$ are the identity losses which is calculated between the real images (CT or CBCT) and the same real image after it has been passed through the opposite generator (CBCT-to-sCT or CT-to-sCBCT respectively) to enforce an identity constraint.

\begin{figure}[h!]
    \centering
    \includegraphics[width=\textwidth]{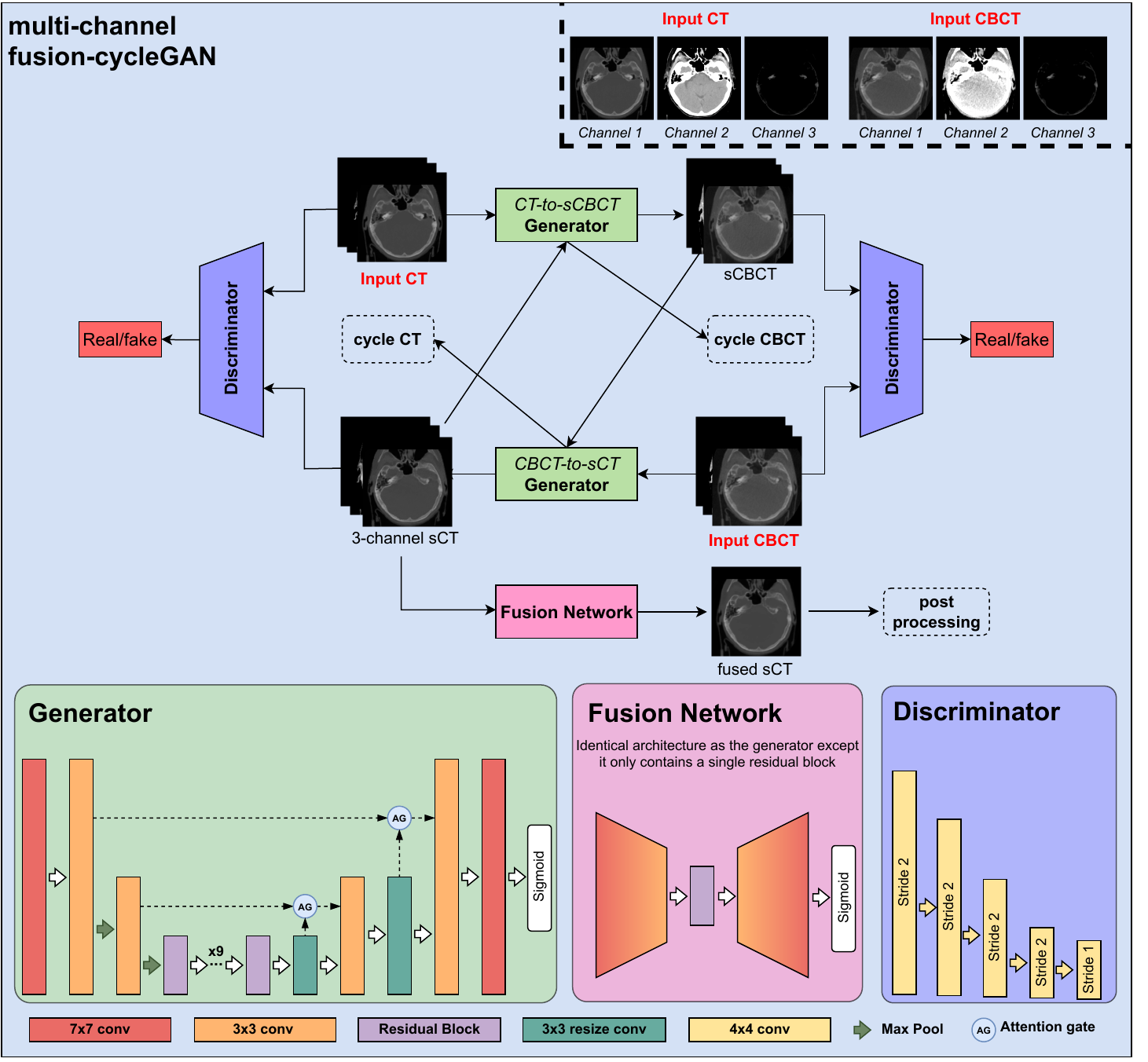}
    \caption{The cycleGAN architecture is used to convert between CBCT and CT image domains, with an additional fusion network following CBCT-to-CT synthesis. Three-channel CBCT and CT images are input into the cycleGAN generators to produce three-channel synthetic CT (sCT) and CBCT output (sCBCT). The fusion network is only applied to the sCT images to recombine channels into a single greyscale sCT.}
    \label{fig:cyclegan}
\end{figure}

\section{Implementation Details}

The model was implemented in PyTorch Lightning version 1.5.10 and two networks were trained independently for each anatomical site. Each model was trained using a single NVidia GeForce RTX 3090 GPU with 24GB VRAM. The training protocols used are shown in Table~\ref{table:training}.

\begin{table}[h!]
\caption{Training protocols.}
\label{table:training}
\begin{center}
\resizebox{0.77\textwidth}{!}{
\begin{tabular}{m{4.5cm}<\raggedright|m{6cm}<\raggedright} 
\hline
Data augmentation & none \\
\hline
Batch size & 1 \\
\hline
Maximum epochs & 200  \\
\hline
Optimizer & Adam   \\ 
\hline
Initial learning rates & \makecell*[l]{Generator: 0.0001\\ Discriminator: 0.0002 } \\ \hline
Learning rate decay schedule & After 5 epochs, decay to $80\%$ of learning rate every 2 epochs for both generator and discriminator. \\
\hline
Stopping criteria, and optimal model selection criteria & Early stopping when the total generator validation loss does not improve for 20 epochs. Optimal model is chosen based on best image similarity metrics calculated on train-time validation data\\
\hline
Loss functions & Mean-squared error on the generators and binary cross-entropy on the discriminators. \\
\hline
Training time & \makecell*[l]{Brain:  $\sim 9$ hours per epoch \\ Pelvis: $\sim 3$ hours per epoch}  \\
\hline
\end{tabular}
}
\end{center}
\end{table}

\subsection{Post-Processing / Channel Fusion}
Upon inference, the trained three-channel CBCT-to-sCT generator network was applied to preprocessed CBCTs (as described in \ref{sec:preprocessing}) to generate a three-channel sCT. After a forward pass of the CBCT-to-sCT generator, the three-channel sCT was input into the fusion network to generate a single-channel fused sCT.

The padding or cropping performed during preprocessing was reversed to restore the sCTs to the original CBCT image size. Next, the HU intensity range of each channel was restored using a linear re-scaling: the fused sCT and channel one of the three-channel sCT were rescaled to [-1024, 3000] HU; channel two was re-scaled to [-100, 100] or [-150, 150] HU for the brain and pelvis, respectively, and channel three was re-scaled to [600, 3000] HU.

Given the three-channel output from our adapted cycleGAN and the single channel from the fusion bolt-on network, two methods of combining the multi-channel information were explored in preliminary testing. Initially, the fusion network output alone was evaluated. Secondly, a post-processing approach was explored to manually combine the three-channel output into a single-channel image. 

This is performed by inserting the second and third (soft tissue and high-density) channels into a full window range output. We tested using both the fused sCT and the first channel sCT from the adapted cycleGAN as the reference full-width intensity range for this approach.

\subsection{Checkpoint Selection}

To select the best model, the top five epochs were saved during training according to the total generator validation loss (Eq.~\ref{eqn:gen_loss}). 

Independently, the validation data is loaded and image similarity metrics (listed in section \ref{sec:evaluation}) are calculated between all slices of the sCT and corresponding CT for each checkpoint. The metrics are averaged across the validation set and ranked according to performance. We select the top-performing model across image similarity metrics.

\section{Evaluation}\label{sec:evaluation}
Full details of the evaluation can be found in \cite{ChallengeDesign2023}. 
Briefly, for the validation phase, the generated sCTs were assessed according to image similarity metrics:
\begin{itemize}
    \item Mean absolute error (MAE)
    \item Structural similarity index (SSIM)
    \item Peak signal-to-noise ratio (PSNR) 
\end{itemize}

The image similarity evaluation was complemented with dosimetric analysis for the test phase. Relative dose difference, dose-volume histogram, and gamma index were used. The dosimetric evaluation was performed for both photon and proton treatment plans.



\section{Results}

\subsection{sCT fusion}

\begin{figure}[ht]
    \centering
    \includegraphics[width=\textwidth]{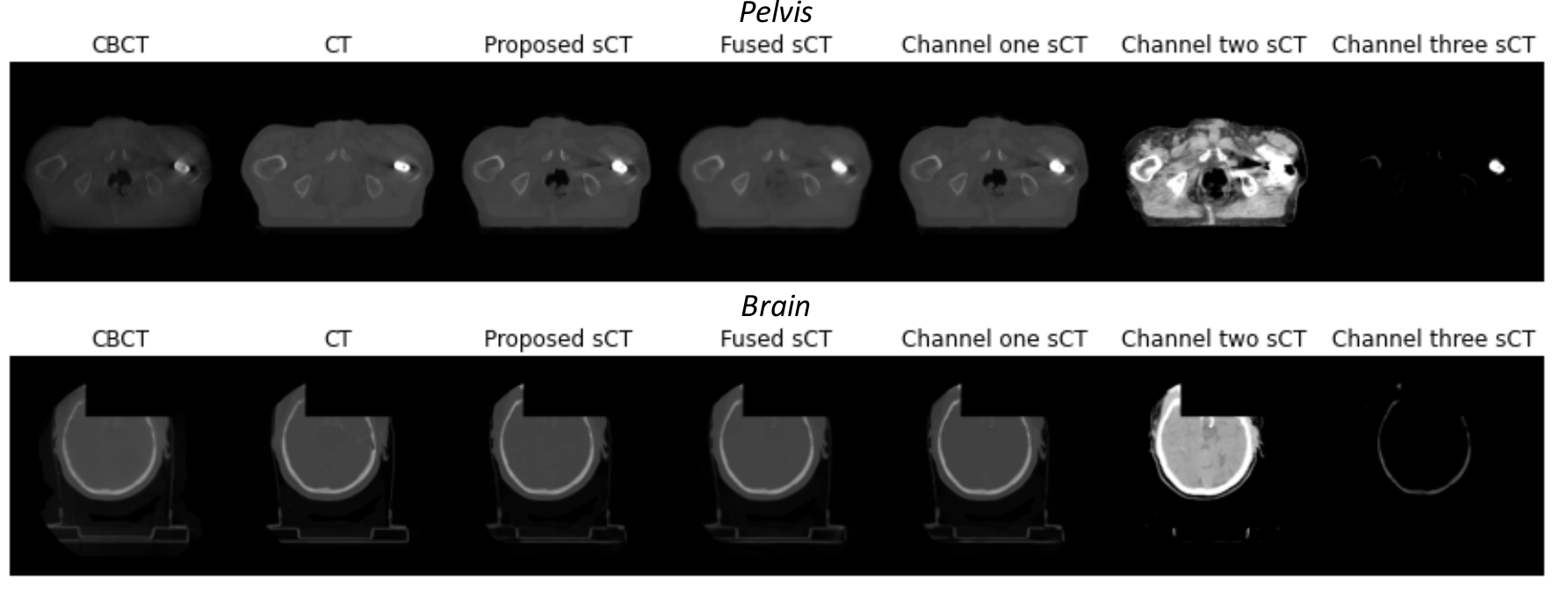}
    \caption{All available synthetic images; the proposed sCT, fused sCT, and individual channel sCTs, with input CBCT and ground truth CT. These images were generated during train-time validation. The first five images are displayed in HU range [-1024, 3000], while channel three range is [600, 3000] and the channel two ranges are [-150, 150] and [-100, 100] for the pelvis and brain, respectively.}
    \label{fig:all_available_sCTs}
\end{figure}

In preliminary testing, we evaluated the direct sCT output from the fusion network. While the direct fused output produced sCTs in a reasonable HU intensity range, the image quality was fairly poor; the soft tissue contrast was subpar and the network failed to reproduce regions of air as shown in Figure \ref{fig:all_available_sCTs}. As a result, the three channels output by the CBCT-to-CT generator were recombined manually. However, as the cycleGAN generator and fusion network were trained end-to-end, we still anticipate that the fusion network may have been a useful meta-learning task which could have aided the model to leverage all three channels by learning their true grey-level relations.

For the pelvis patients, we used channel one of the three-channel output of our model (channel one sCT) as the full range image and reinserted the second and third channels. For the brain, we used the output from the fusion network (fused sCT) as our base image.

\subsection{Results for challenge validation set}

\begin{figure}[ht!]
    \centering
    \includegraphics[width=\textwidth]{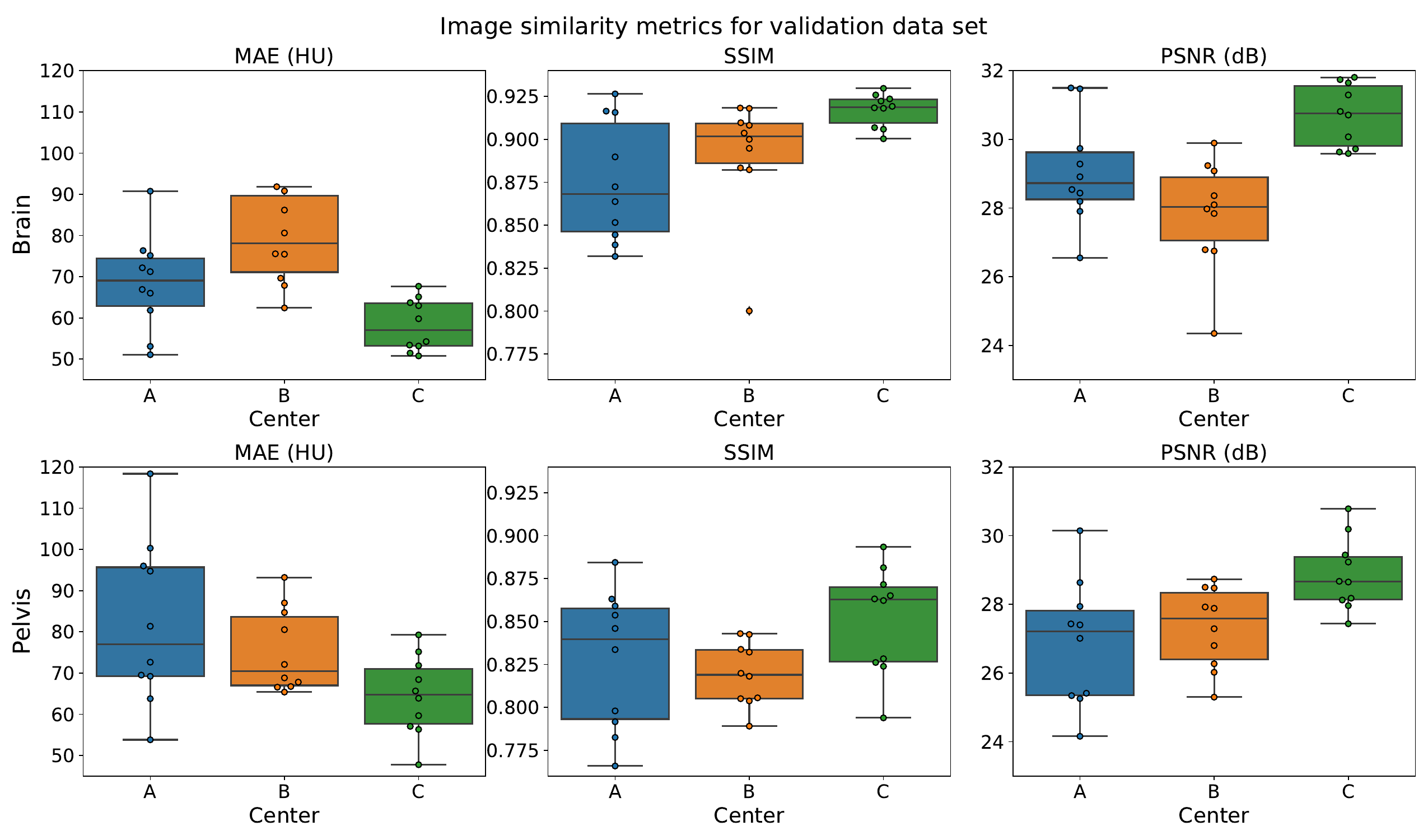}
    \caption{Image similarity metrics; mean absolute error (MAE), structural similarity index measure (SSIM) and peak signal-noise ratio (PSNR) calculated between ground truth CT and sCTs generated from CBCTs in the validation dataset.}
    \label{fig:boxplots}
\end{figure}

Table \ref{tab:metrics} and Figure \ref{fig:boxplots} summarise the quantitative results obtained by image similarity metrics; MAE, PSNR and SSIM. For both anatomical sites, our proposed methods achieve the highest performance on patients from center C, with reduced MAE and increased PSNR and SSIM. Whilst all centers suffer from outliers, our model performs least consistently for patients from center A.

\begin{table}
\centering
\setlength\tabcolsep{1.6mm}
\caption{\label{tab:metrics}Quantitative comparison between the sCTs and ground truth, the pCT.} 


\begin{tabular}{@{}*{7}{c}{c}}
\hline                              
\textbf{Anatomical Site} &\textbf{MAE (HU)}&\textbf{PSNR (dB)}&\textbf{SSIM}\cr 
\hline
Brain (Validation) &$69.70 \pm 15.54 $&$29.20  \pm 1.77 $&$0.89 \pm 0.03 $ \cr
\hline
Pelvis (Validation) & $73.91 \pm 15.13 $&	$27.69  \pm 1.57 $ &	 $0.83 \pm 0.03$  \cr 
\hline
 Full data set (Validation) & $71.83 \pm 15.00 $&	$28.44  \pm 1.85 $ &	 $0.86 \pm 0.05$  \cr
 \hline
\end{tabular}

\end{table}

\begin{figure}[h!]
    \centering
    \includegraphics[width=\textwidth]{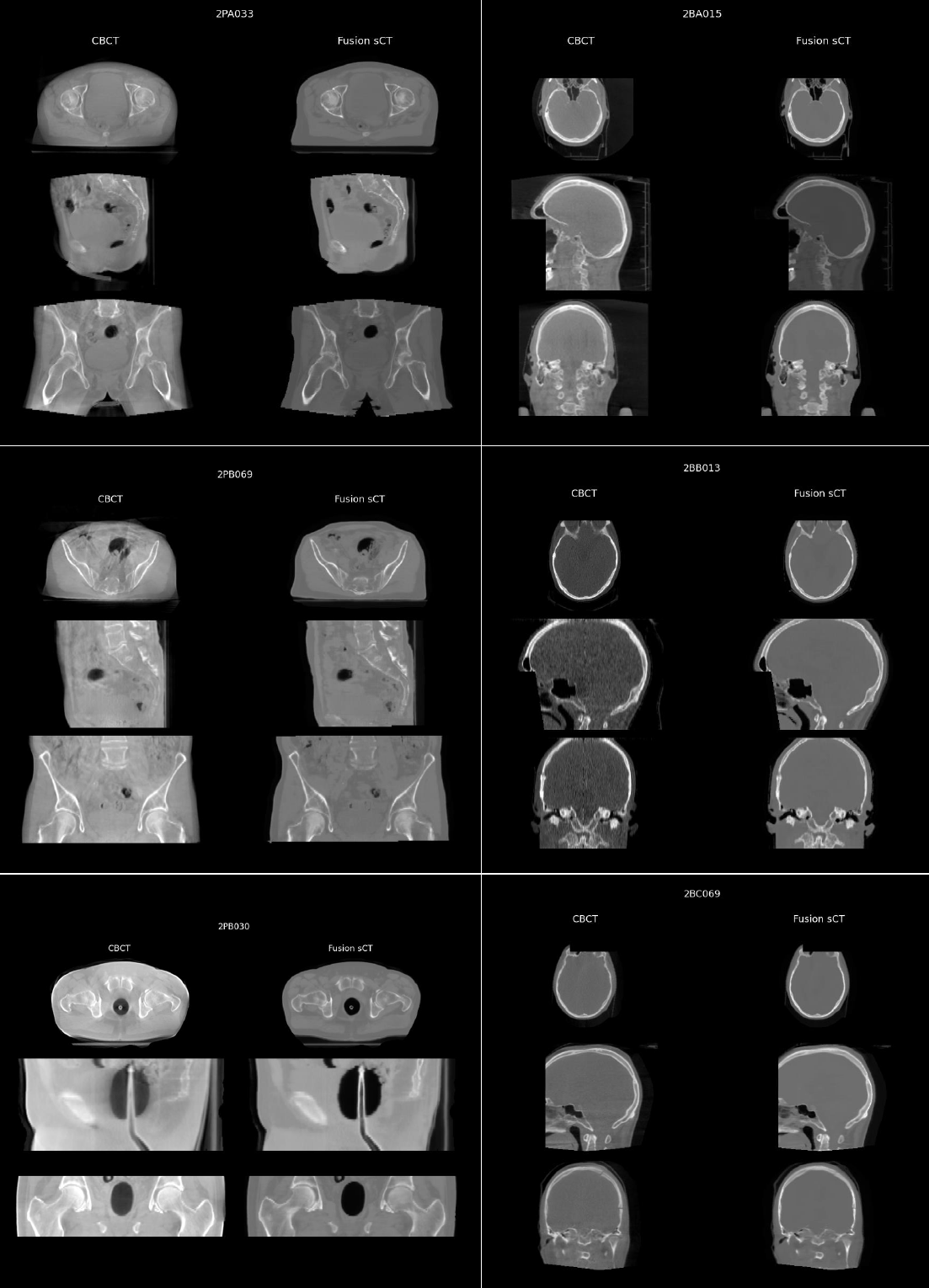}
    \caption{CBCT and sCT generated by our methods. Pelvis and brain results are shown with HU scale of [-400,1200] and  [250, 2500], respectively.}
    \label{fig:output}
\end{figure}

The sCTs generated by our method are displayed in Figure \ref{fig:output} for a subset of patients for each anatomical site from the validation dataset alongside the original CBCT. Visually, the methods preserve the anatomical structure of the CBCT images and the contrast in soft tissue is enhanced for both pelvis and brain patients. For the brain, high-quality sCTs are generated despite the significant decrease in the quality of CBCTs from center B. 

\noindent Observed on the CBCT for pelvis patient 2PB069 in Figure \ref{fig:output} are significant artefacts on the axial CBCT slice that have been considerably suppressed by our method. Whilst we see a large reduction in streaking artefacts, improving visualisation of internal anatomy, the artefacts appear to have influenced the sCT body contour. This is also noticed for patient 2PA033. 

\begin{figure}[h!]
    \centering
    \includegraphics[width=0.67\textwidth]{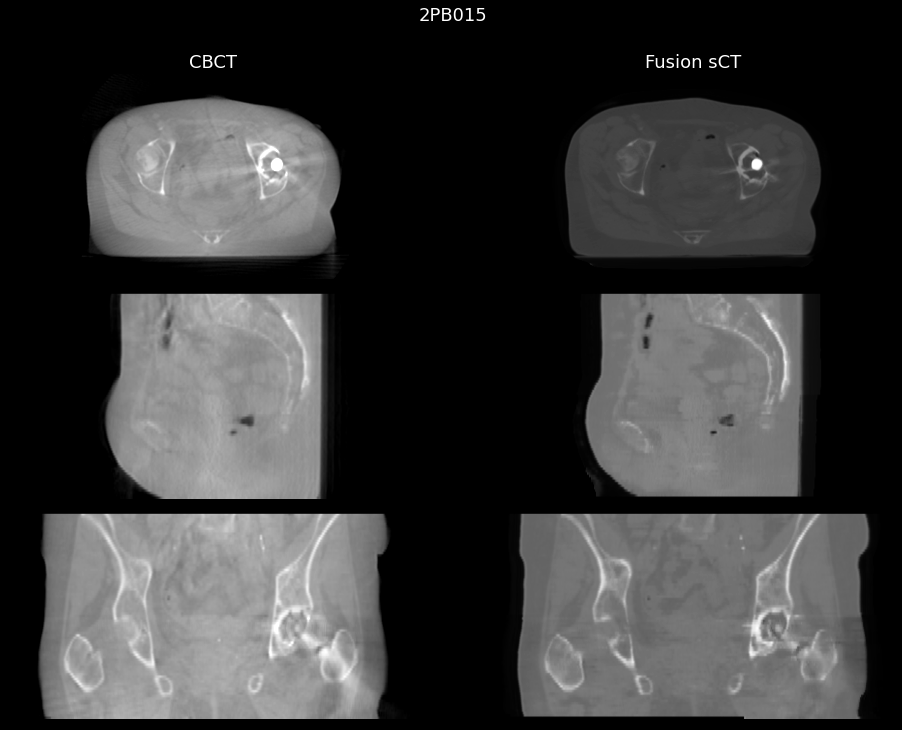}
    \caption{CBCT and sCT generated by our methods for example pelvis patient with metal implant in the hip. Images are shown with HU scale of [-400,1200] and  [250, 2500], respectively.}
    \label{fig:output_metal}
\end{figure}

Our method is also robust to scatter artefacts caused by high-density features such as metal hip implants. The CBCT and sCT for a pelvis patient with an artificial hip are shown in Figure \ref{fig:output_metal}, demonstrating accurate preservation of the metal and reduction in streaking in the nearby tissue.

\subsection{Results for challenge test set}

Tables \ref{tab:test_metrics} and \ref{tab:dose_metrics} provide a summary of the quantitative results derived from both image similarity and dose evaluation metrics, calculated on the test set provided by the challenge organisers. Our methods demonstrate consistent and commendable performance on this test dataset, showing marginal improvements across the average image similarity metrics. 

\begin{table}
\centering
\setlength\tabcolsep{1.6mm}
\caption{\label{tab:test_metrics}Quantitative image comparison between the sCTs and ground truth, the pCT, calculated by challenge organisers for test data set.} 

\begin{tabular}{@{}*{7}{c}{c}}
\hline                              
\textbf{Anatomical Site} &\textbf{MAE (HU)}&\textbf{PSNR (dB)}&\textbf{SSIM} \cr 
\hline
 Full data set (Test) & $71.58 \pm 13.79  $&	$28.34  \pm 1.50 $ &	 $0.86 \pm 0.04$  \cr
 \hline
\end{tabular}
\end{table}

Our methods demonstrate good accuracy in dose evaluation for photon intensity-modulated treatment plans, with a gamma index of $98.42\%$ when using a $2\%$ dose-difference criteria with $2mm$ distance-to-agreement criteria. However, this accuracy is diminished when applied to proton plans. This is expected due to the heightened sensitivity of proton doses to small anatomical variations.

\begin{table}
\centering
\setlength\tabcolsep{1.6mm}
\caption{\label{tab:dose_metrics}Quantitative dose comparison between the sCTs and ground truth, the pCT, calculated by challenge organisers for the test data set.} 

\begin{tabular}{@{}*{7}{c}{c}}
\hline                              
\textbf{Plan} &\textbf{DVH}&\textbf{Gamma Index (\%) }&\textbf{Dose MAE} \cr 
\hline
Photon & $0.07 \pm 0.02  $&	$98.42  \pm 4.95 $ &	 $0.01 \pm 0.01$  \cr
Proton & $0.27 \pm 0.27  $&	$92.32  \pm 5.87 $ &	 $0.07 \pm 0.04$  \cr
 \hline
\end{tabular}
\end{table}



%
%
\section{Discussion and Conclusion}

Our approach demonstrates the successful generation of high-quality synthetic CT images from CBCT scans. The incorporation of image data at varying windows/levels within a concatenated multi-channel input has allowed us to effectively address several challenges inherent in CBCT imaging. We observe notable improvements in the preservation of fine soft tissue details, suppression of artefacts such as streaking, and accurate representation of high-density features within the patient anatomy. 
Our reported image similarity metrics on the validation dataset show an average MAE of $71.83 \pm 15.00 $ HU, PSNR of $28.44 \pm 1.85 $ dB, and SSIM of $0.86 \pm 0.05$. Similarly, on the test dataset, the average MAE is $71.58 \pm 13.79 $ HU, PSNR is $28.34 \pm 1.50 $ dB, and SSIM is $0.86 \pm 0.04$. These results indicate a high degree of similarity to the ground truth images, which are CT scans registered via non-rigid image registration to the input CBCTs.

Based on current validation metrics, we have observed that sCTs generated from center C outperform those from both center A and B for both brain and pelvis patients, despite center C having a reduced number of slices (approximately half) in the pelvis training set. Specifically, the MAE is $64.51 \pm 9.11$ HU for center C, compared to $82.21 \pm 18.71$ HU and $75.19 \pm 9.62$ HU for centers A and B, respectively. This outcome suggests that our model generalises effectively between centers, and we attribute the discrepancy partially to the fact that images from center C appear cropped, particularly in regions where higher errors are observed for other centers - the end slices. Since these slices lack clinical significance, the performance of center C is more indicative of our model effectiveness.


%


We chose to exploit a multi-channel input to capture specific information about both the CBCT and CT images. The first channel spans the full image range for simplicity upon training, while the second channel focuses on narrow window widths to enhance soft tissue contrast. For high-intensity features and dense bone, the third channel is employed. Our investigations into training with different channel widths indicate a sensitivity of learned features to window width, necessitating further research into the optimal window/level for each channel as well as channel quantity.

Our fusion methods differ between brain and pelvis sites. In the case of brain images, the fused sCT can be used in the recombination of the three channels. However, it negatively affects the quality of the sCT images for pelvis patients. The fusion network architecture is modelled after the generator network in the cycleGAN. In this setup, the output is compared to a CT image registered only through rigid image registration. Since pelvis patients' anatomy can change unpredictably over even short time intervals, the semi-paired nature of our training could be inappropriate. On the other hand, brain patients experience fewer such changes, allowing the fusion network to perform well despite the absence of deformable image registration between the CBCT and CT images. While cycleGANs are designed to operate in an unsupervised manner with unpaired data, there is an expectation that incorporating non-rigid image registration into the training process could lead to an overall performance improvement. Our future efforts will concentrate on refining the fusion process, including optimizing the design of the fusion network architecture

CBCT to sCT image synthesis holds the potential to enhance visualization and improve dose calculations for adaptive radiotherapy \cite{Rusanov2022}. Our method succeeds in generating high-quality synthetic CT images from on-treatment CBCTs. We demonstrate that by incorporating multiple window/level inputs and effectively fusing them, we can enhance the quality of the resulting sCT images and achieve good dose accuracy.

\subsubsection{Acknowledgements} The authors of this paper declare that the method implemented for participation in the \textit{SynthRAD2023} challenge has not used any pre-trained models nor additional datasets other than those provided by the organisers. This work would not have been possible without the support of the radiotherapy-related research group in Manchester. The authors would like to give special thanks to Marcel van Herk, Alan McWilliam, Eliana Vasquez Osorio, Andrew Green, Robert Chuter and Jane Shortall for their support and guidance. CS is funded by Elekta AB.

%
%
%
\bibliographystyle{splncs04}
%

\bibliography{References}
\end{document}